\documentclass[12pt]{article}

\usepackage{graphics}

\begin{document}

\title{A Method of Areas for Manipulating the Entanglement Properties of
One Copy of a Two-Particle Pure Entangled State}

\author{Lucien Hardy \\
Centre for Quantum Computation, The Clarendon Laboratory,\\
Parks road, Oxford OX1 3PU, UK}

\maketitle

\begin{abstract} We consider the problem of how to manipulate the
entanglement properties of a general
two-particle pure state, shared between Alice and Bob,
by using only local operations at each end and classical
communication between Alice and Bob.  A method is developed in which
this type of problem is found to be equivalent to a problem involving
the cutting and pasting of certain shapes along with certain colouring
problems.  We consider two problems.  Firstly, we find the most general
way of manipulating the state to obtain maximally entangled states.
After such a manipulation the entangled states $|11\rangle +|22\rangle
+\dots +|mm\rangle$ are obtained with probability $p_m$.  We obtain an
expression for the optimal average entanglement obtainable.  Also, some results of
Lo and Popescu pertaining to this problem are given simple geometric
proofs.  Secondly, we consider how to manipulate one two-particle
entangled state $|\psi\rangle$ to another $|\psi'\rangle$ with
certainty.  We derive Nielsen's theorem (which states a necessary and
sufficient condition for this to be possible) using the method of areas.
\end{abstract}

\vspace{6mm}

\section{Introduction}

Quantum entanglement has many applications including quantum
teleportation \cite{tele}, quantum cryptography \cite{cryp}, and
quantum communication \cite {comm}. This has
led people to regard entanglement as a resource.  However, entanglement
can exist in different forms and so it is useful to know how it can
be manipulated from one form to another.  In this paper the problem of
manipulating a general pure two-particle entangled state in order to
obtain maximally entangled states will be considered.  We will also
consider the problem of how to manipulate one general pure two-particle
entangled state to
another.  Alice and Bob are allowed to do whatever they
want locally and they are allowed to communicate classically with one another.
They are not allowed to exchange quantum states.  This type of situation has
already been much
discussed in the literature.  The problem of how to optimally manipulate
a large number, $N$, of copies of a general pure two-particle entangled
states into maximally entangled states by local means has been
completely solved in the asymptotic limit
$N\rightarrow\infty$ \cite{concen}.  However, the perhaps more basic problem of how to
manipulate a
single copy of a general pure two-particle state into maximally
entangled states has not been so extensively discussed.  The most
significant work on this is by Lo and
Popescu \cite{lopop} who prove certain bounds relating to this problem.  However
their proofs, while being extraordinarily ingenious, are rather difficult
to follow.  The method developed in this paper, which completely solves
the problem, involves the cutting and pasting of areas along with a
colouring problem. Once
the basic methods have been put in place, it is very easy to picture
what is happening.  This method is used to find the maximum obtainable average
entanglement and to derive a formula of Lo and Popescu which gives the
maximum probability of obtaining a given maximally entangled state.

A related problem is how to transform one pure two-particle entangled
state to another, and to establish which states will transform to one
another in this way.  Nielsen \cite{nielsen} has completely solved this
problem. However his proof of his theorem uses some unfamiliar mathematics.
An alternative reasonably simple proof of Nielsen's theorem is given
here which, again, involves cutting and pasting of areas along with a certain
colouring problem.

\section{Obtaining maximally entangled states}

\subsection{Introduction}

The most general pure two-particle state can be written in the Schmidt
form
\begin{equation}\label{psi}
|\psi\rangle= \sum_{i=1}^I \sqrt{\lambda_i} |i\rangle_A|i\rangle_B
\end{equation}
where we choose $\lambda_i \geq \lambda_{i+1}$ and where the states
$|i\rangle_{A,B}$ are orthonormal.  We want to manipulate this state in
order to obtain states which are of the form
\begin{equation}\label{mstate}
|\varphi_m\rangle= {1\over\sqrt{m}}\sum_{k=1}^m |k\rangle_A|k\rangle_B
\end{equation}
We will call this state an $m$-state.  An $m$-state is equivalent to
$\log_2m $ copies of 2-states \cite{concen}. After the process is completed we
should have a certain $m$-state with a certain probability $p_m$.
Particle A goes
to Alice and particle B goes to Bob.  Alice and Bob are allowed to perform
whatever operations they want locally and also they communicate classically
with each
other.  This can happen in the following way. Alice performs a
measurement and communicates the
result to Bob who then performs a measurement which depends on the
result of Alice's measurement, and then Bob communicates his result back
to Alice and she makes another measurement, and so on back and forth.
This is most general way in which Alice and Bob can manipulate their
state without actually exchanging quantum states in the process.
Whatever measure of entanglement we employ, the amount of
entanglement should not increase during such a process.  Lo and Popescu
show that, for the very special case of a two-particle pure state, this process
is equivalent to one in which Alice makes one measurement and
communicates the result to Bob who then may perform a unitary evolution
operation on his particles.   The reason for this significant
simplification is that, due to the Schmidt decomposition, any operation
by Bob is equivalent, so far as the resulting form of the state is
concerned, to some operation by Alice.  Hence, Alice can
simply do everything herself in one go and then communicate the final
result to Bob.  The most general measurement Alice can make is a POVM.
This is equivalent to Alice introducing an ancilla, $S$, performing a
general unitary evolution on particle $A$ and the ancilla $S$, and then
making a projective measurement on the ancilla. Let us imagine that the
ancilla has
a basis set of states $|l\rangle_S$. Since we allow completely general
evolution of $A+S$ we can assume, without loss of generality, that the
final measurement projects onto subspaces spanned by the states
$|l\rangle_S$.  Furthermore,
it is shown in the appendix that there is no advantage to be had by
performing a non-maximal (i.e. degenerate) measurement and so we can
assume that this
measurement is maximal and projects onto the operators
$|l\rangle_S\langle l|$. For each
outcome, $l$, the two particles $A$ and $B$ should be projected into an
$m$-state $|\varphi_{m_l}\rangle^l$ where
\begin{equation}
|\varphi_{m_l}\rangle^l= {1\over\sqrt{m_l}}\sum_{k=1}^{m_l}
|k\rangle_A|k\rangle_B^l
\end{equation}
The superscript $l$ is included since we do not require that the Schmidt
vectors satisfy $|k\rangle_{B}^l=|k\rangle_{B}^{l'}$ for
$l\not=l'$.  After Alice has communicated the result, $l$, of the
measurement to Bob, Bob could rotate these
vectors into the same standard form for all $l$ (thus removing the need
for the superscript at this stage), but this is not
important.  It is enough that Alice and Bob know what $l$ is so
they know what state they have.  We do not require a superscript $l$ on
the $|k\rangle_A$ states since, as explained below, Alice can
rotate her Schmidt vectors to
standard form as part of the overall unitary transformation she
performs.  Just before Alice makes her
measurement projecting onto $|l\rangle_S$ the state of the system will be
\begin{equation}\label{targeta}
|\Psi_{\rm target}\rangle= \sum_l \sqrt{\mu_l}|l\rangle_S
|\varphi_{m_l}\rangle^l
\end{equation}
where the coefficients $\sqrt{\mu_l}$ can be taken to be real since
any phases can be absorbed into appropriately redefined $|l\rangle_S$.
Note that if, at this stage, the $|i\rangle_A$ states had a superscript
$l$ then they could be rotated into standard form by applying a series of
controlled unitary operations, $\hat U_l$, to $A$ where the control is
the $|l\rangle_S$ state.  At this stage Alice has not done anything which
is irreversible. Having completed her local manipulations Alice will
perform a maximal projective
measurement.  It follows from the result of Lo and Popescu that
manipulating the state into the form (\ref{targeta}) and then measuring
onto the $|l\rangle_S$ basis is equivalent to the most general procedure
for manipulating the two particles by local means to $m$-states.  We
will use equation (\ref{targeta}) later when we come to show that the
method developed in the next section is equivalent to the most general
method.

\subsection{How to obtain maximally entangled states}

\begin{figure}
\resizebox{3.5in}{!}
{\includegraphics{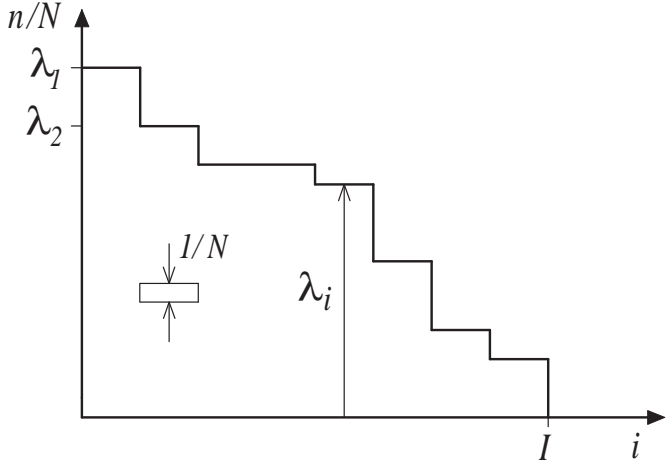}}
\caption{The start state can be plotted on a graph.  Such a graph is
called an area diagram.  This area diagram has step structure in which
the steps go down towards the right.}\label{fig1}
\end{figure}

Now consider the initial state $|\psi\rangle$ given in equation
(\ref{psi}).  We will introduce an ancilla,  $R$, in the state
$|1\rangle_R$.  This ancilla has basis states $|n\rangle_R$ where $n=1, 2,
\dots N$.  We will take $N$ to be very large and will want to consider the
case where $N$ tends to infinity.  We define the integers
$N_i=N\lambda_i$ where, for the moment, we are taking the $\lambda_i$ to
be rational numbers so the integers $N_i$ can be found with $N$ finite.
This constraint can be relaxed when $N$ tends to infinity.
The initial state of $RAB$ is $|1\rangle_R|\psi\rangle$.  Alice now
evolves $R+A$ using the following transformations.
\begin{equation}\label{firsttrans}
  |1\rangle_R|i\rangle_A \rightarrow {1\over\sqrt{N_i}}(\sum_{n=1}^{N_i}
  |n\rangle_R)|i\rangle_A
\end{equation}
The fact that these transformations evolve orthogonal states to orthogonal
states ensures that they can be implemented by unitary evolution.
Under these transformations the state $|1\rangle_R|\psi\rangle$ evolves to
\begin{equation}\label{eqamp}
|\Psi_{\rm start}\rangle= {1\over\sqrt{N}} \sum_i \sum_{n=1}^{N_i}
|n\rangle_R|i\rangle_A|i\rangle_B
\end{equation}
which we will call the {\it start state}.
We see that each of the $N$ terms in this superposition has the same
amplitude.  Each of these terms will be represented by a rectangular
element width 1
and height $1/N$.  The area of the element is $1/N$ and equal to the
probability associated with the corresponding term in (\ref{eqamp}).
Each element can be labelled by $(n,i)$ corresponding to the term
$|n\rangle_R|i\rangle_A|j\rangle_B$ (initially $i=j$ but after Alice has
performed operations on her particles this will not necessarily be the
case for every term).  The elements are then arranged on a graph where
elements having the same $n$ are placed in the same row and elements
having the same $i$ are placed in the same column.  The resulting graph
looks like a series of steps as shown in Fig. \ref{fig1}.
We will call this the
{\it area diagram}.  The total area under the steps is 1 corresponding
to the total probability.  On the vertical axis $n/N$ is plotted.  On the
horizontal axis the $i$ which appears in $|i\rangle_A$ is plotted.  For small
$n$ every $i$ position will be filled.  However, because of the form of
the state (\ref{eqamp}), once $n$ gets bigger than $N_I$ there will
no longer be any $|I\rangle_A|I\rangle_B$ terms.  This is the reason for
the step at
$n=N_I$.  There will be further steps at each $n=N_i$.  The height of
the rightmost step is $N_I/N=\lambda_I$.  Subsequent steps will be at
heights $\lambda_i$ as shown in Fig. \ref{fig1}.  If a projective measurement
were to be performed on $R$ at this stage then, corresponding to
each outcome $n$, the state of $A+B$ would be projected onto an $m$-state
where $m$ is equal to the number of elements in the $n$th row on the area
diagram. This will give a distribution of $m$-states with probabilities
which are equal to the large horizontal rectangular areas of width $m$
and height $\lambda_m-\lambda_{m+1}$ formed by extending the
horizontal parts of the step back to the vertical axis.  However, the
area in the diagram can be moved around by Alice in a way to be
described below by performing local
unitary operations.  When this is followed by a projective measurement on $R$
different distributions of $m$-states can be realised.

The terms $|n\rangle_R|i\rangle_A|j\rangle_B$ and
$|n'\rangle_R|i'\rangle_A|j'\rangle_B$ are bi-orthogonal iff
${}_R\langle n|n'\rangle_R {}_A\langle i|i'\rangle_A=0$ (orthogonal at
Alice's end) and ${}_B\langle j|j'\rangle_B=0$ (orthogonal at Bob's
end).  If two terms are only orthogonal at either Alice's end or at
Bob's end then they are mono-orthogonal.
In the area diagram we will impose the constraint that all
elements in a row, that is for a given $n$, correspond to terms which
are bi-orthogonal.  This
ensures that when we perform a projective measurement onto $R$ the
resulting state will be an $m$-state.

We will colour all the elements which correspond to terms which are
mono-orthogonal to one another a given colour.  Elements corresponding
to terms which are
bi-orthogonal will be coloured with different colours.  Thus, initially,
all the elements in a given column are the same colour and every column
is coloured a different colour to every other column.  When area is
moved around the constraint that terms corresponding to elements in a
row be bi-orthogonal means that all elements in any given row must be
different colours.

The method of areas to be developed here
involves moving area elements around in such a way that there is a net
movement of area up and to the left.   We will see that
it is possible to have a net movement of area up the area diagram but
not down the diagram. This means that the net movement of area across
any horizontal line drawn on the diagram must be up.
The basic unitary operation,
$U(n,i \leftrightarrow n',i')$, employed by Alice is
defined by the transformation equations
\begin{equation}\label{swap}
 |n\rangle_R|i\rangle_A  \rightarrow |n'\rangle_R|i'\rangle_A
\end{equation}
\begin{equation}
 |n'\rangle_R|i'\rangle_A  \rightarrow |n\rangle_R|i\rangle_A
\end{equation}
with no change for all other $|n''\rangle_R|i''\rangle_A$.
We will call this the swap operation.
The effect of this operation is to move elements around on the area
diagram.  If there are elements at both the $(n,i)$ and $(n',i')$
positions then they will have their positions swapped.  If, there is
only an element at one of the two positions then it will be moved to the
other position while the original position will become vacant.  These
moves will not effect Bob's part of the state.  If two terms are
bi-(mono)-orthogonal before the swap operation is applied to one or both of
them then they will be bi-(mono)-orthogonal afterwards.  In other words,
the swap operation does not change the colour of the elements.
Initially, as stated above, all elements in the same row are different
colours and elements in the same column are the same colour.
In moving elements of area around we impose
only the constraint that, at the end of the process, all elements in a row
are different colours.  Although elements in any given column start off
being the same colour, we do not demand that this is true at the end of
the process.  We can move a large number of elements at once.
In the limit as $N \rightarrow \infty$ the
elements will become infinitesimal in height.  Hence, in this limit, we can
make horizontal cuts anywhere.  We can make vertical cuts along the
edges of the columns.  The area can be cut up into smaller pieces and
then pieces can be moved around and pasted into new positions.  It is
possible to move area around like this in any way we want by repeated
applications of the swap operation.  The empty space above the steps can
be used as a clipboard for the temporary storage of pieces of area to
facilitate the rearrangement of area if required.

\begin{figure}
\resizebox{4in}{!}
{\includegraphics{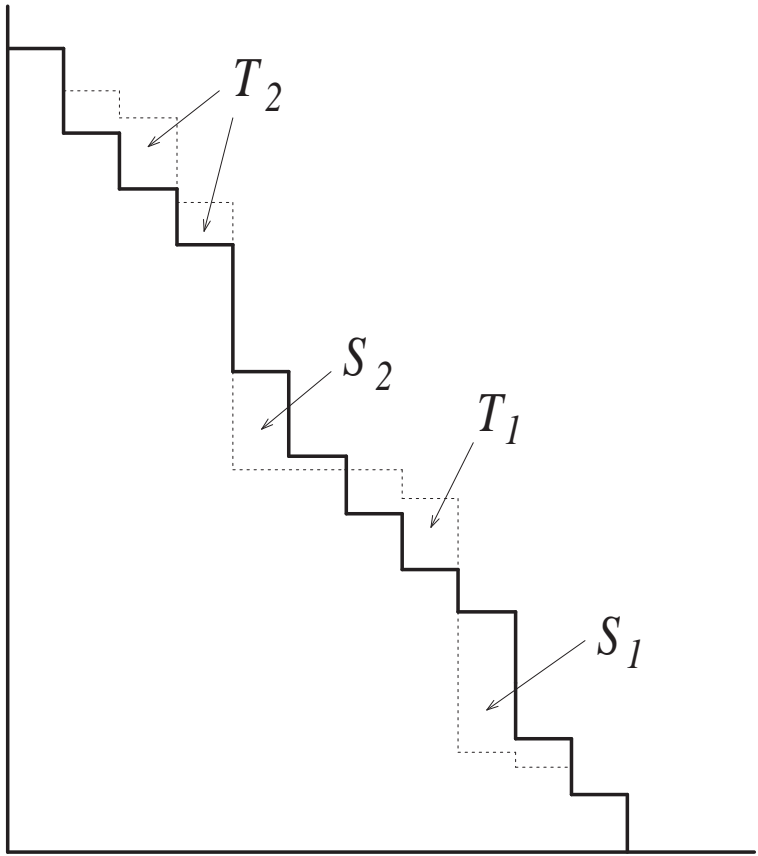}}
\caption{
Area is transferred up the area diagram to go from the start steps shown
by the bold line to the target steps shown by the dashed line.  The
total area transferred up is $S=T$
}\label{fig2}
\end{figure}

In Fig. \ref{fig2} the original step structure is shown by a full line.  A new
step structure is shown by a dashed line.  We will impose the constraint
(to be justified later) that the new step structure consists of steps
which, like the original steps, go  across and down (but never up) towards the
right.  The area $S=\sum S_r$ which will be
cut away from some parts of the steps is equal to the area $T=\sum T_r$
which will be added to other parts of the steps.
Each of these areas consist of $R$ smaller disjoint parts labelled by $r$
going up the
diagram.  Note, area $T_r$ lies between area $S_r$ an $S_{r+1}$.  Note
also that the areas $S_r$ and $T_r$ can themselves be made up of
disjoint parts.
The constraint that there is no net movement of area downwards means
that $\sum_{r=1}^{r'} T_r \leq \sum_{r=1}^{r'}S_r$ for all $r'$.
The original steps are of height $\lambda_i$.  Let the new steps be of
height $\lambda'_i$. Since the columns are of unit width, these lengths
are numerically equal to the areas of the columns, and hence the
constraint that area is moved only to the left
is equivalent to the set of constraints
\begin{equation}\label{newsteps}
\sum_{i=p}^I \lambda'_i \leq \sum_{i=p}^I \lambda_i
\end{equation}
for all $r=1$ to $I$ with equality holding when $r=1$.

We could simply move the area $S$ to
the area $T$ element by element by applying the swap operator.  However,
if we did this it is likely that elements corresponding to
mono-orthogonal terms would end up in the same row.
If we can redistribute the area so that it corresponds to
the new step structure but without there being any elements of the same colour
in the same row then we will have realised another distribution of
$m$-states.  This is because Alice could then perform a projective
measurement on $R$ and corresponding to each outcome, $n$, will be a
$m$-state and these $m$-states will clearly have a different
distribution.  The probability of a given $m$ state is equal to the
area of the horizontal rectangle formed by projecting leftwards the top and
bottom of the step at position $m$.  This rectangle has width $m$ and
height $\lambda'_m-\lambda'_{m+1}$.  Hence, for the new area diagram,
the probability of getting an $m$ state is
\begin{equation}\label{psubm}
p_m=(\lambda'_m-\lambda'_{m+1})m
\end{equation}
We will show firstly that this colouring problem can be
solved and secondly that the process described here is general in the
sense that any distribution of $m$-states which can be achieved by local
means can be achieved by this method.  Hence, equations (\ref{newsteps})
and (\ref{psubm}) define the possible distributions of $m$-states that
can be obtained.

\begin{figure}
\resizebox{2.5in}{!}
{\includegraphics{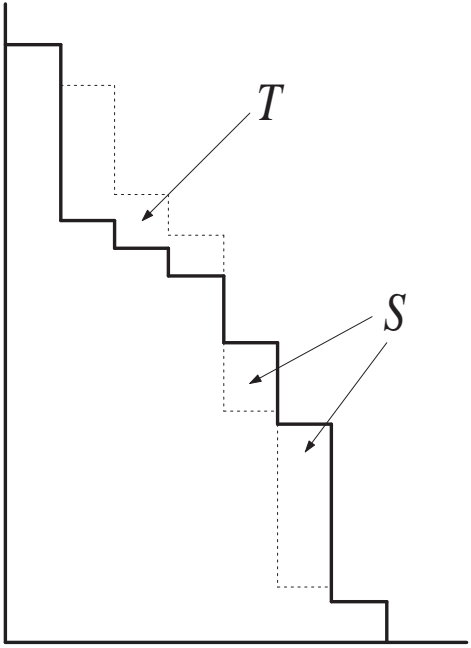}}
\caption{
This figure shows the start steps (bold line) and the target
steps(dashed line) of the recolouring problem discussed in the text.
}\label{fig3}
\end{figure}

\begin{figure}
\resizebox{\textwidth}{!}
{\includegraphics{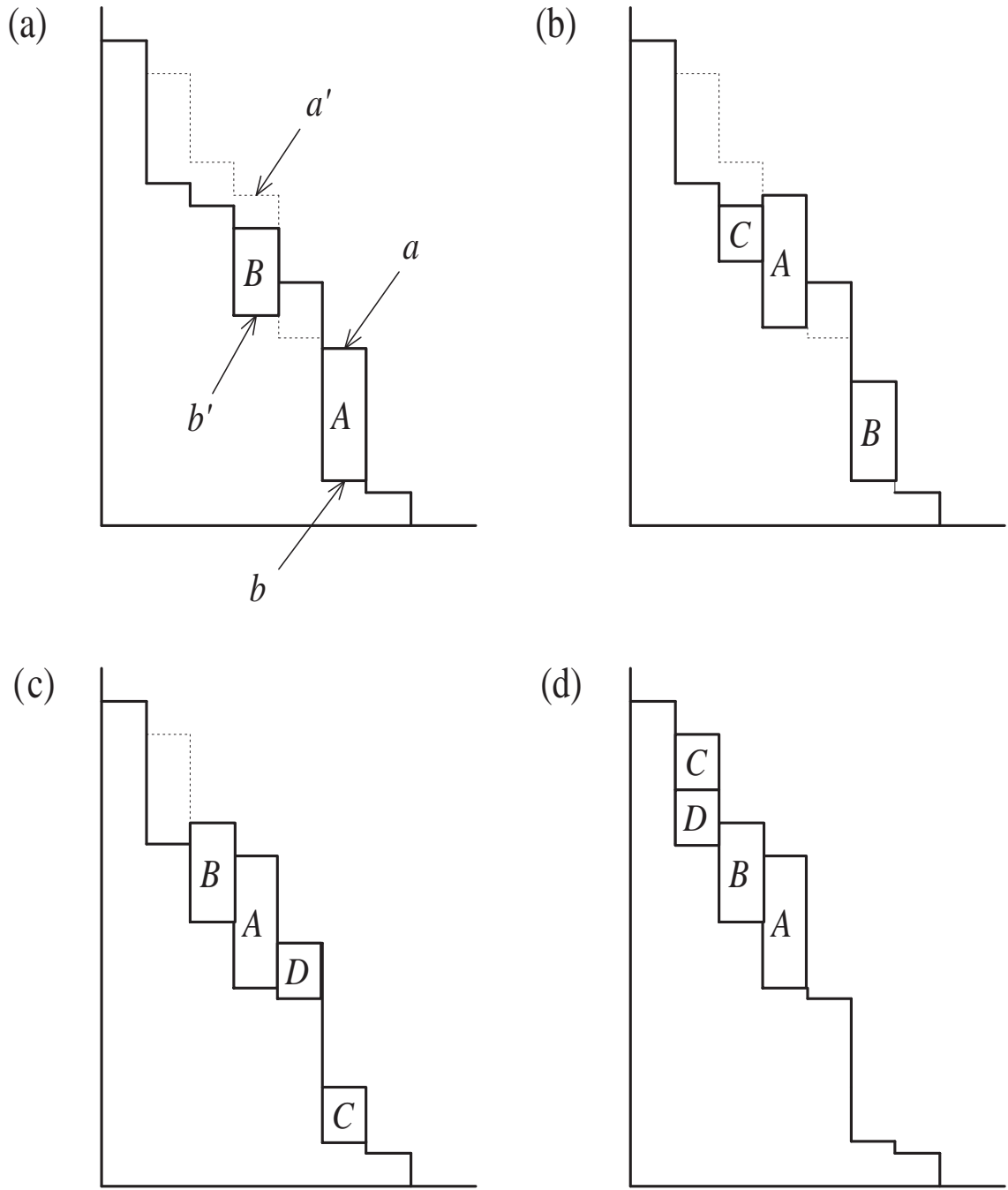}}
\caption{
This figure shows the recolouring procedure described in the text.
After the recolouring, no colour is in more than one place across any
given row.
}\label{fig4}
\end{figure}

The solution to the colouring problem will be explained by reference to
the example shown in Fig. \ref{fig3}.
This example has $R=1$ (since all of $T$
is above all of $S$).  However, it will be clear that the method
works for the general case.  We start by taking the rightmost column in
the area $S$.  This is area $A$ in Fig \ref{fig4}(a). This area is then swapped
into the rightmost column of the area $T$ so that point $a$ coincides with
point $a'$ as shown in Fig. \ref{fig4}(b).
However, it may be too long
to all fit in to the right most column of area $T$ (as is
indeed the case in our example), and hence
it will displace some of the area in the column below this, marked as
area $B$ in Fig. \ref{fig4}(a). Because of the nature of the swapping operation,
this area $B$ will be moved to the old position of $A$ so that point $b$
coincides with point $b'$ as shown in Fig. \ref{fig4}(b).  Area $A$ is now in its
final position.
Let us imagine that the column from which area $A$ was taken was
coloured red.  This red colour will now be divided between
what is left of the original column and the area $A$ in its new
position.  Since the steps go up towards the left, it is impossible to
have red at more than one position in any given row. The rightmost
column of $T$ is now filled so we start on the next rightmost column.
We swap area
$B$ into position in the next rightmost column of area $T$.  Again, this
could be too long.  In our example it is too long and projects into
the column below into the
area marked $C$.  Hence $C$ will be moved to where $B$ was. This takes
us to the situation shown in Fig. \ref{fig4}(c). Now we move area $C$ into position
in the next rightmost column of area $T$.  This could again be too long
and project into the column below, but in this example this
is not the case.  Rather, area $C$ is too short leaving a gap.  Hence no
area is moved back to the rightmost column of $S$ and this means we have
finally dispensed with the net effect of moving the area from this
rightmost column of $S$.  Now we select the next rightmost column of
$S$.  In our example this area is marked $D$ in Fig. \ref{fig4}(c).
This area is
now moved up following the same procedure.  Thus, we move $D$ into
position as high as possible in the rightmost column of $T$ which has
not yet been filled.
This places it below area $C$. It could be the case that $C$ is too long
and projects into the column below in which case some area would be
swapped back and we would have to continue as before.  However, in our
particular example area $D$ fits in below $C$ and the recolouring is
finally completed as shown in Fig. \ref{fig4}(d). This method can be applied to
any recolouring
problem of this nature.  The general method is that area is swapped
from the rightmost non-empty column of $S$ to as high a position as
possible in the rightmost non-full
column of $T$.  We note that a given colour can end up in, at most, two
different columns and that any colour moved from a column will always
end up higher than its original position.  This means that it is
impossible for two
elements of area which started in the same column (having the same
colour) to end up in the same row.  Hence, the colouring problem has
been solved.

\subsection{Proof that this is most general method}

Having shown that it is possible to have a net movement of area up the
area diagram in any
general way which is consistent with maintaining the step structure,
we will now show that this corresponds to the most
general way of manipulating entanglement to produce $m$-states in the sense
that any distribution of $m$-states which can be achieved by local operations
classical communication can be achieved by this method.
The idea of this proof will be to show that the target state (just
before Alice measures on to the ancilla) can be put into a certain form
which is inconsistent with any movement of area downwards.

We have already established that the most general final state just
before Alice makes her measurement (we call this the {\it target state}) is
the state given in (\ref{targeta}).
If we write $\nu_l=\mu_l/m_l$ then (\ref{targeta}) becomes
\begin{equation}\label{targetap}
|\Psi_{\rm target}\rangle=   \sum_l \sum_{k=1}^{m_l}
\sqrt{\nu_l}|l\rangle_S|k\rangle_A|k\rangle_B^l
\end{equation}
We can consider further unitary transformations by Alice on this state
to put it into a form in which every term has the same amplitude.  Let
the dimension of the ancilla $S$ be
$M$ and define $M_l=M\nu_l$ (again we will let $M\rightarrow\infty$) and
let Alice perform the following transformations on (\ref{targeta}):
\begin{equation}\label{transab}
  |l\rangle_S \rightarrow
  {1\over\sqrt{M_l}}\sum_{n\in W_l} |n\rangle_S
\end{equation}
where $W_l$ is the set of $M_l$ integers from $(\sum_{r=0}^{l-1} M_r)+1$
to $\sum_{r=0}^l M_r$ (we put $M_0=0$).  Under this transformation
(\ref{targeta}) becomes
\begin{equation}\label{targetb}
|\Psi'_{\rm target}\rangle= {1\over\sqrt{M}}\sum_l
\sum_{k=1}^{m_l}\sum_{n\in W_l}
    |n\rangle_S|k\rangle_A|k\rangle_B^l
\end{equation}
Every term in this state has the same amplitude.  In arriving at this
state from the previous target state (\ref{targeta},\ref{targetap})
we have done nothing that is
irreversible.  Furthermore, if we measure onto the $|n\rangle_S$ basis we
are just as likely to get a given $m$-state as with the previous target
state.  Hence, we can regard this as our new target state.  Any
method by which $m$-states can be obtained is equivalent to manipulating
the state into the form (\ref{targetb}).

If we project onto $|n\rangle_S$ we will obtain an $m_n$-state where $m_n$
can be read off from (\ref{targetb}).  We can relabel the $n$'s such
that $m_{n+1}\leq m_n$.  Thus, we can impose the following:
\begin{description}
\item[Constraint A] If outcome $n$ corresponds to a $m_n$-state then, without
loss of generality, we can impose the constraint that $m_{n+1}\leq m_n$.
\end{description}
By examining (\ref{targetb}) we can see that there is a second
constraint that can be imposed on the form of the final target state.
\begin{description}
\item[Constraint B]  For the target state we can, without any loss of
generality, impose the constraint that
\begin{equation}\label{conb}
|{}_B\langle \theta|{}_S\langle n|\Psi'_{\rm target}\rangle|^2 \leq {1\over M}
\end{equation}
where $|\theta\rangle_B$ is any normalised state for system $B$ since
this is true of equation (\ref{targetb}).
\end{description}

We can identify the ancilla $S$ with the ancilla $R$ introduced earlier.
Hence, $S\equiv R$ and $M=N$.  Now consider the state
(\ref{psi}).  Since Alice's operations do not effect Bob's
system we see that we have the following constraint:
\begin{description}
\item[Constraint C] While the state $|\Psi\rangle$ is being manipulated by
local unitary operations by Alice, we will always have
\begin{equation}\label{conc}
|{}_B \langle j|\Psi\rangle|^2 = \lambda_j={N_i\over N}
\end{equation}
for all $j$.
\end{description}
Since (\ref{eqamp}) is related to (\ref{psi}) by reversible operations,
we can take (\ref{eqamp}), which corresponds to an area diagram, as our
starting point.  We will now see that
area cannot be moved down in the area diagram.  For the purposes of this
proof consider a change in the way the area diagram is plotted such that
the $j$ in $|j\rangle_B$ (rather than the $i$ in $|i\rangle_A$) is plotted
on the horizontal axis.  This will simply have the effect of
redistributing elements
horizontally but not vertically since $n/N$ is still plotted on the
vertical axis.  We label elements in this modified area diagram by
$\{ n,j\}$.   Constraint C implies that the column
corresponding to a given $j$ on this modified area diagram will always
have the same area.  This is because Alice's actions cannot effect the
total area (or probability) associated with column $j$.  However, Alice
can change the $n$ value associated with area elements and hence she can
move area in column $j$ up and down.  It is possible that she can bring
about a net movement of area (or probability) down in this column.  This
would lead to the area being compressed into a smaller space than it
would \lq\lq naturally\rq\rq ~fit.  Any net movement of area downwards,
whether on the $\{ n,j\}$ picture or the $(n,i)$ picture would
correspond to this happening in at least one $j$ column.  This is
exactly what we want to rule out.  We will now see that any such net
movement downwards will violate constraint $B$ (which we were free to
impose on the target state).

If, at some stage, the state has been manipulated to the
state $|\Psi\rangle$ then the area of the $\{ n,j\}$ element will be
$A_{nj}=|{}_B\langle j|{}_S\langle n|\Psi\rangle|^2$.  Initially, for the
start state $|\Psi_{\rm start}\rangle$ in (\ref{eqamp}), we have
\begin{equation}
\sum_{n=1}^{n'} A_{nj}^{\rm start}={n'\over N}
\end{equation}
for all $n'\leq N_i$.
However, if there is a net movement of area down the area diagram with
respect to the horizontal line $n=n'$ then, since
the total area in a given column is constant (by constraint C), this net
movement of area
downwards must happen in at least one column of the modified area
diagram.  Hence, for at least one
value of $j$, we must have
\begin{equation}\label{areabig}
\sum_{n=1}^{n'\leq N_i} A_{nj}^{\rm target}>{n'\over N}
\end{equation}
However, since $M=N$, equation
(\ref{conb}) implies
\begin{equation}\label{areasmall}
\sum_{n=1}^{n'\leq N_i} A_{nj}^{\rm target}\leq{n'\over N}
\end{equation}
The contradiction between (\ref{areabig}) and (\ref{areasmall}) proves
that a net movement of area downwards is not possible on the modified
area diagram and hence neither is it possible on
the unmodified area diagram.  Note that this proof goes through for any
sort of operations by Alice and in particular it does not assume that
the only operations Alice can make are the swap operations defined in
(\ref{swap}).

To complete the proof that the manipulations described earlier are
equivalent to the most general way of manipulating the state to obtain
$m$-states we note the following.
\begin{description}
\item (i) The target state (\ref{targetb}) can be represented on an area
diagram in which $n/N$ (where $N=M$) is on the vertical axis and $m_n$
is plotted on the horizontal axis.  Since $m_n$ is an integer and since
we can impose constraint A without loss of generality, this area diagram
will have a step structure (in which the steps of integer width
go down towards the right).
\item (ii) The initial state can be taken to be the start state
$|\Psi_{\rm start}\rangle$ given in (\ref{eqamp}) and this can be
represented by an area diagram with the step structure.
\item (iii) The total area of both these diagrams is 1.  Therefore the most
general way of manipulating the start state into $m$-states
corresponds to
going from one area diagram with the step structure to another with the
step structure in a way consistent with the constraint that there is no
net movement of area downwards.
\item (iv) The method, employing the swap operator, discussed previously can
be used to go from one step structure to another in any way that is
consistent with there being no net downward movement of area.  Hence, it is
equivalent to the most general method.
\end{description}

\subsection{Getting the highest possible average}

If we
are only interested in the average amount of entanglement in the form of
maximally entangled states we can obtain,
this being equal to $E=\sum_m p_m \log_2 m$, then it turns out
that any movement of area will decrease this average.
Hence this average has a maximum given by the original area diagram
\begin{equation}\label{emax}
E^{\max} = \sum_m (\lambda_{m}-\lambda_{m+1})m \log_2 m
\end{equation}
To see that any movement of area will decrease this average, consider
moving one element on the area diagram (with area equal to ${1\over N}$)
from the end of row $A$ which has original width $m_A$, to the end of row
$B$ which has original width $m_B$.
Since we can only move area to the left we require that $m_B +1 < m_A$
(if $m_B+1=m_A$ then the rows will simply have interchanged their lengths
and hence there will be no change in the distribution of $m$-states).  The
original contribution of these two rows to $E$ will be
\begin{equation}
\Delta E_{\rm initial}={m_A\over N} \log_2 m_A +{m_B \over N}\log_2 m_B
\end{equation}
The contribution afterwards will be
\begin{equation}
\Delta E_{\rm final}=\left({m_A\over N} -{1\over N}\right) \log_2 (m_A-1)
+\left({m_B \over N}+{1\over N}\right)\log_2 (m_B+1)
\end{equation}
The difference between these two contributions is
\begin{equation}
\Delta E_{\rm final}-\Delta E_{\rm initial}
= {1\over N} \log_2 \left[
    \left({m_B+1\over m_B}\right)^{m_B} \left( {m_A-1\over m_A}\right)^{m_A}
    \left( { m_B+1 \over m_A-1}\right)  \right]
\end{equation}
It can only be advantageous to move elements of area if this quantity
is positive.  However, by making the substitutions
\begin{equation}
m_A=x+{1\over 2} + r  \qquad m_B=x-{1\over 2}-r
\end{equation}
we can see that $\Delta E_{\rm final}-\Delta E_{\rm initial}$ is
negative if $r > 0$.  The constraint that $m_A > m_B+1$ implies that
$r>0$ and hence any movement of area must lead to a smaller $E$.
We also see from this that since only one distribution of $m$-states
leads to the maximum $E$, any attempt to alter the distribution of
$m$-states will result in a decrease of $E$ (and so we obtain another
main result of Lo and Popescu).

\subsection{Proof of a formula of Lo and Popescu}

\begin{figure}
\resizebox{4in}{!}
{\includegraphics{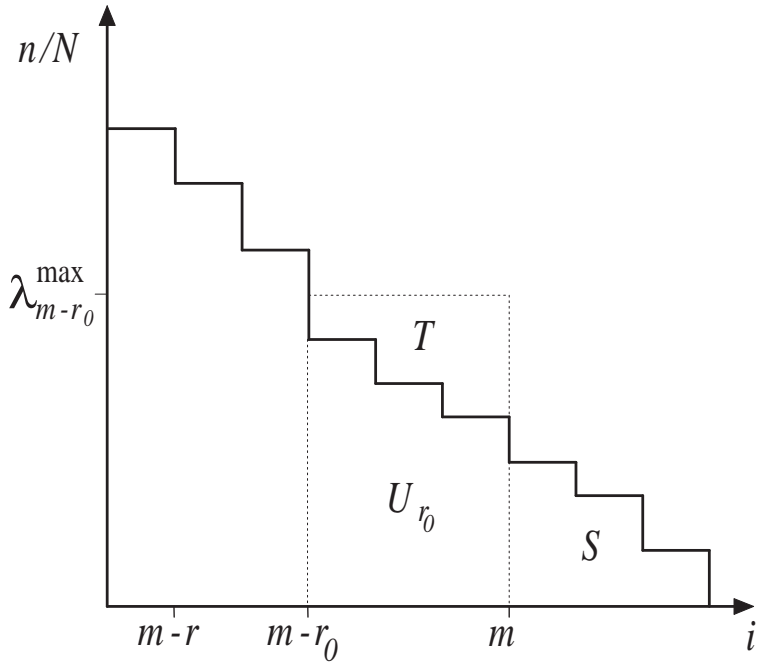}}
\caption{
This figure shows a strategy for obtaining the maximum probability of a
given $m$-state.  The area $T$ is equal to the area $S$.  This defines
$r_0$ and $\lambda_{m-r_0}^{\rm max}$
}\label{fig5}
\end{figure}

We will now use this method to derive a formula central to the paper of
Lo and Popescu \cite{lopop}.  Imagine
that we have a general two-particle pure entangled state and we want to
have a given $m$-state with as high a probability as possible.  We want
to know what this probability is and what strategy to use.
This corresponds to the area redistribution shown in Fig. \ref{fig5}.
The target area diagram consists of a block of width
$m$ with an additional bit on top.  This defines $r_0$ and
$\lambda_{m-r_0}^{\rm max}$ (see diagram).
The area $S$ has been moved to the area $T$ where these two areas are equal.
The height of the main block
is $\lambda_{m-r_0}^{\rm max}$ which can be calculated  since we know
that
\begin{equation}\label{rzero}
r_0 \lambda_{m-r_0}^{\rm max}=U_{r_0}+T=U_{r_0}+S=\sum_{i=m-r_0+1}^I
\lambda_i
\end{equation}
where the last equality follows from the fact that the $i$th column is of
area $\lambda_i$.  The total probability of getting the $m$-state is
equal to the area of the main block, i.e. $p_m^{\rm max}=
m \lambda_{m-r_0}^{\rm max}$.
We can use this formula if we know what $r_0$ is since then
$\lambda_{m-r_0}^{\rm max}$ can be calculated from (\ref{rzero}).  This can be
established by the following considerations.  Define the area $U_r$ to
consist of
all the columns $i=m-r+1$ to $m$ in the start area diagram so that
\begin{equation}
U_r=\sum_{i=m-r+1}^m \lambda_i
\end{equation}
where $r=1,2\dots m$.  The area $U_r+T$ consists of a main
block of width $r$ and height $\lambda_{m-r_0}^{\rm max}$ plus, for
$r\not=r_0$, an extra
bit lying outside this block (when $r< r_0$ a bit of $T$ lies outside
this block and when $r>r_0$ a bit of $U_r$ lies outside the block).
Hence,
\begin{equation}
U_r + T \geq r \lambda_{m-r_0}^{\rm max}
\end{equation}
with equality in the case $r=r_0$.  Therefore,
\begin{equation}
\lambda_{m-r_0}^{\rm max} = \min_r [{1\over r} (U_r +T)]
\end{equation}
and we obtain the formula of Lo and Popescu
\begin{equation}
p_m^{\rm max}= \min_r {m\over r} (\sum_{i=m-r+1}^I \lambda_i)
\end{equation}
where $r=1, 2, \dots m$.  The geometric origin of this formula is now
clear.

\section{Proof of Nielsen's Theorem}

\subsection{Introduction}

The set of constraints (\ref{newsteps}) are exactly Nielsen's condition
\cite{nielsen} for
being able to manipulate an entangled state, $|\psi\rangle$,
with Schmidt coefficients $\sqrt{\lambda_i}$ to another, $|\psi'\rangle$, with
Schmidt coefficients $\sqrt{\lambda'_i}$.  However, we cannot
immediately interpret the new area diagram as being equivalent to a
$|\psi'\rangle$ state since, unlike in the original area diagram, a
given column can be multicoloured.  We will see that, nevertheless,
we can use the area diagrams to prove Nielsen's theorem.  This proof
works along similar lines to the previous proof.  First, we put the
target state into step form.  Alice introduces an ancilla $S$ of
dimension $M$ with
basis states $|l\rangle_S$.  If the problem can be solved then, for similar
reasons to before, she must be
able to manipulate the total state into the form
\begin{equation}\label{targetg}
|\Psi_{\rm target}\rangle=\sum_l {\sqrt{\mu_l}} |l\rangle_S |\psi'\rangle^l
\end{equation}
where
\begin{equation}\label{prime}
|\psi'\rangle^l= \sum_{i=1}^I \sqrt{\lambda'_i}|i\rangle_A|i\rangle_B^l
\end{equation}
Substituting (\ref{targetg}) into (\ref{prime}) we obtain
\begin{equation}\label{ntarget}
|\Psi_{\rm target}\rangle= \sum_l\sum_i \sqrt{\mu_l\lambda_i}
|l\rangle_S|i\rangle_A|i\rangle_B^l
\end{equation}
Define $M_{li}=\mu_l\lambda_i M$.  We apply the transformation
\begin{equation}
|l\rangle_S|i\rangle_i \rightarrow
{1\over\sqrt{M_{li}}} \sum_{n\in W_{li}} |n\rangle_S|i\rangle_A
\end{equation}
where $W_{li}$ is the set of integers from $(\sum_{k=0}^{l-1}
M_{ki})+1$ to $\sum_{k=0}^l M_{ki}$ (we set $M_{0i}=0$).  Under this
transformation (\ref{ntarget}) becomes
\begin{equation}\label{nbtarget}
|\Psi'_{\rm target}\rangle = {1\over\sqrt{M}}
\sum_l\sum_i\sum_{n\in W_{li}} |n\rangle_S|i\rangle_A|i\rangle_B^l
\end{equation}
Each term has equal amplitude in this state.
If we project it onto $|n\rangle_S$ then we will get a state with,
say, $m_n$ terms (where $m_n$ can be read off from (\ref{nbtarget})) not
all of which are bi-orthogonal.  We can relabel the $n$'s so that
$m_n\geq m_{n+1}$ and hence we can draw an area diagram with a step
structure.  The terms corresponding to a given $n$ are not necessarily
bi-orthogonal and hence we cannot impose on the target area diagram the
constraint
that elements in a row must all be coloured different colours.  Rather we
will have a different colouring problem.  As before, we will identify the
systems $S$ and $R$ so that $M=N$.

\subsection{Obtaining Nielsen's bound}

\begin{figure}
\resizebox{4in}{!}
{\includegraphics{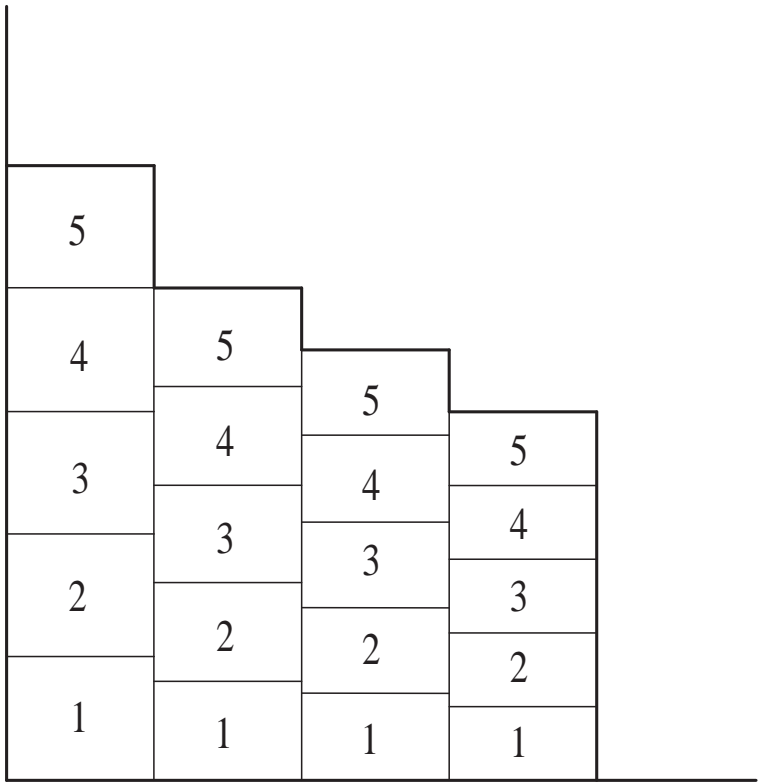}}
\caption{
The columns of the target diagram are divided up into $Q$ equal parts.
In this figure we have $Q=5$ but we will let $Q\rightarrow\infty$
}\label{fig6}
\end{figure}

The start state is $|\Psi_{\rm start}\rangle$ in (\ref{eqamp}) and is
represented by an area diagram with step structure in which each column
is coloured with only one colour, this being different to the colour of
the other columns.  The target area diagram is
shown in Fig. \ref{fig6}. Since this has been recoloured the columns in this
diagram will not necessarily be of one colour.
The height of the $i$th column is $\lambda'_i=N'_i/N$.  We
chose a number $Q$ which we will let tend to infinity, though in the
Fig. \ref{fig6} we have set $Q=5$.  We divide each column up equally into $Q$
pieces which are numbered $q=1, 2, \dots Q$ starting at the bottom.
The $q$th piece in the $i$th column is labelled $[q,i]$
If the target area diagram has been obtained from the start area diagram
by moving finite sized bits of area
around (as will be the case in our recolouring strategy) then there
will be a finite number of horizontal boundaries
between different colours.  Some of the pieces $[q,i]$ are likely to
have these boundaries on them.  However, as $Q\rightarrow\infty$ the
total area of such pieces will tend to zero and we can assume that each
piece has a unique colour.  The idea will be to collect all the pieces
with a given $q$.  Since their areas are proportional to $\lambda'_i$
they can correspond to the new state $|\psi'\rangle^q$.  However, each of
these pieces having the same $q$ must be coloured with a different
colour since the terms in $|\psi'\rangle^q$ are bi-orthogonal.  There
are $N'_i$ terms in the state
vector corresponding to each column, and since these are divided into
$Q$ pieces, there are $N'_i/Q$ terms corresponding to the $[q,i]$ piece
which will be of the form (unnormalised)
\begin{equation}
|[q,i]\rangle = {1\over\sqrt{N}} \sum_{n=(q-1)N'_i/Q+1}^{qN'_i/Q}
|n\rangle_R|i\rangle_A|j_{qi}\rangle_B
\end{equation}
The total state is the sum of all such terms.  We can transform the
total state by applying the transformation
\begin{equation}
{\sqrt{Q\over N'_i}}\sum_{n=(q-1)N'_i/Q+1}^{qN'_i/Q}|n\rangle_R|i\rangle_A
\rightarrow |q\rangle_R|i\rangle_A
\end{equation}
for all $q, i$.  This sends the term $|[q,i]\rangle$ to the state
\begin{equation}
|[q,i]'\rangle = \sqrt{\lambda'_i \over Q}
|q\rangle_R|i\rangle_A|j_{qi}\rangle_B
\end{equation}
This transformation has simplified the state of the ancilla $R$
for the terms corresponding to each piece $[q,i]$.  In so doing we
have recovered the coefficient $\sqrt{\lambda'_i \over Q}$.  The total
state is now
\begin{equation}
|\Psi'\rangle =  {1\over\sqrt{Q}}\sum_{q=1}^I
\sqrt{\lambda'_i}|q\rangle_R|i\rangle_A|j_{qi}\rangle_B
\end{equation}
Now, if we measure onto the $|q\rangle_R$ basis we get a state which
is a realisation of $|\psi'\rangle$ iff the terms are bi-orthogonal.  In
colouring terms, this means we require that all the pieces labelled
with the same $q$ in Fig. \ref{fig6} should be of different colour. Thus, we have
another colouring problem. If we can solve this colouring problem under
the assumption that net movement of area is up, then
we will have given a constructive proof that Nielsen's bound can be
obtained.  To complete the proof of Nielsen's theorem we need to show
that area can only be moved up.  This will be done later.

\begin{figure}
\resizebox{\textwidth}{!}
{\includegraphics{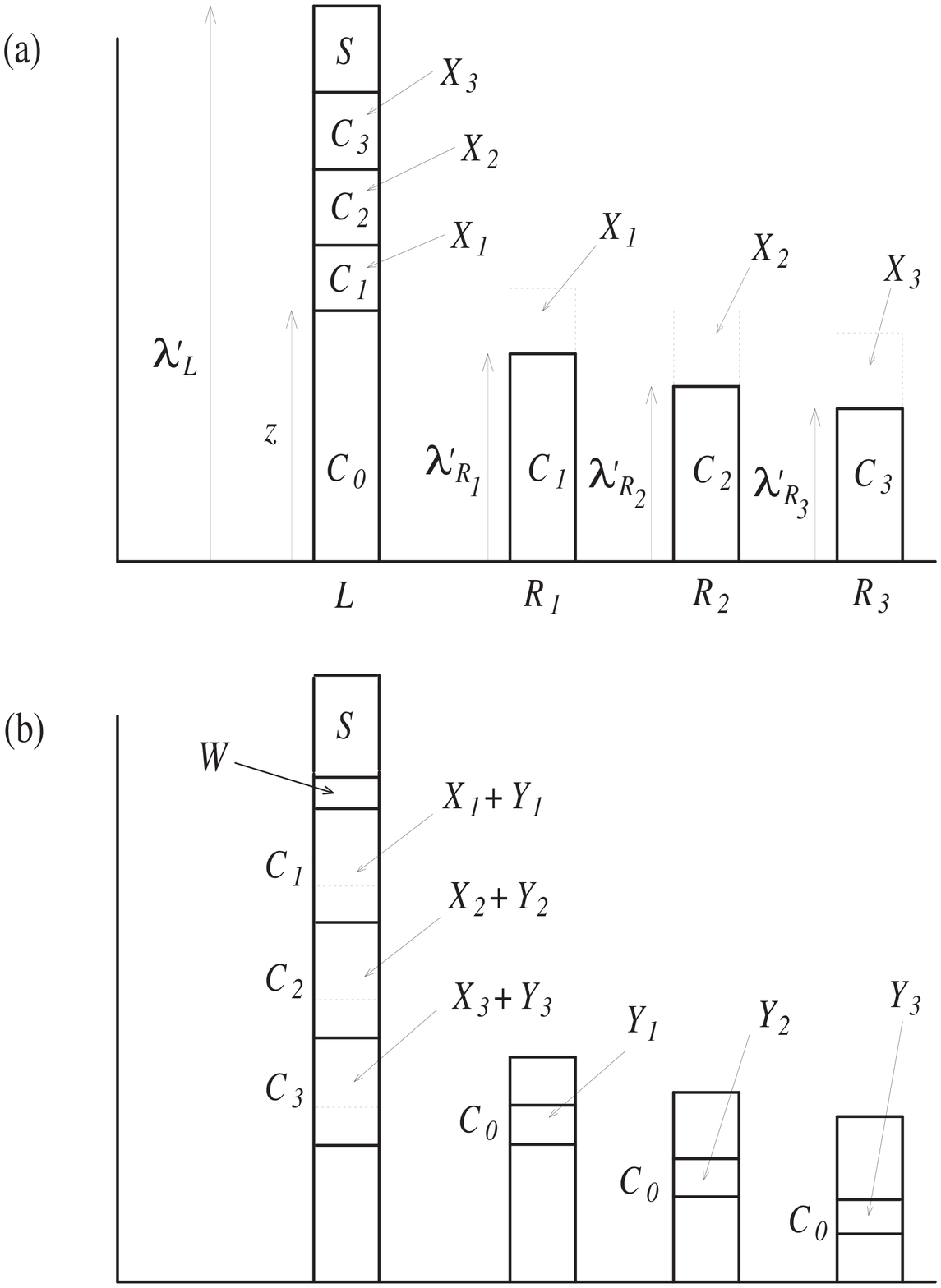}}
\caption{
To solve the colouring problem we consider correct bunches of columns
such as those shown here.  We
correct the diagram by moving the areas $Y_k$
}\label{fig7}
\end{figure}

A way to solve this colouring problem was
suggested to the author by A. Mahtani.  This solution is obtained simply
by correcting the solution to the previous colouring problem in
Sec. 2.2 \cite{othersoln}
Firstly, this colouring procedure, illustrated by example in Fig.
\ref{fig4},
is used to go from the start area diagram (representing
$|\psi\rangle$) to the target diagram where the heights of the columns
are $\lambda'_i$.  Now we note that it is a
property of this recolouring procedure that a given colour
can end up in two columns at most.  There are two ways in which a piece
of area can be moved to the left.  Either it can be moved directly (as
are the pieces $A$ and $D$ in Fig. \ref{fig4}, or it can first be
swapped to the
right and then be moved back to a further left position than it started
in (as are pieces $B$ and $C$ in Fig. \ref{fig4}).  We will call
the first type {\it directly swapped} pieces and the second type
of pieces the {\it swapped back} pieces.  Since these swapped back
pieces (for example, piece $B$) must be shorter than the piece that
displaced them from their original column (in the case of $B$ this was
piece $A$), since they always end up at the top of their destination
column, and since the column they end up in is higher than their
starting column, they must occupy proportionally less of their final
column than the pieces that displaced them (piece $B$ occupies
proportionally less of its final column than piece $A$ of its final
column).  This means that when the columns are divided up into a large
number, $Q$, of equal pieces the colours of the swapped back pieces will
not appear in two pieces with the same $q$. Hence, we will first correct
the other colours.  However, the procedure which corrects the colours
of the directly swapped pieces disturbs this property of the swapped
back pieces.  Hence, after correcting for a bunch of directly swapped
pieces, we will have to correct for the swapped back pieces as well.
First we will consider the colours corresponding to the directly swapped
pieces.  Before we do that note that some columns will remain unchanged
(neither have area moved into them, or out of them) and hence their
colour cannot end up in two pieces with the same $q$.  These columns can
be completely ignored and we can
consider only the remaining columns.  We start at the rightmost column
and go left considering only the columns which have changed.  The first
changed columns we meet will be shorter than originally, will each be
coloured with only one colour, and will have had some area swapped out
of them directly.
After one or more of these monochromatic columns, we will meet a column into
which these directly swapped pieces have been moved.  The last
directly swapped piece may displace a piece from this column which will
be a swapped back piece $S$.  This piece, when it is swapped back,
will end up at a column, $L$,
somewhere to the left.  But before we get there, we may meet a few more
monochromatic columns, $R_k$, from which pieces $X_k$ have been directly
swapped. We will count $k=K,K-1,\dots 1$ backwards so first we meet
$R_K$ as we go
leftwards.  Eventually, on our journey leftwards, we meet column $L$ into
which $S$ has been swapped, then pieces $X_K$ to $X_1$.  When the last
piece $X_1$ is swapped into position it may displace piece $S'$ which
will be swapped back. Thus, we will repeat the same story as we continue
leftwards.  Columns will come in bunches of a few monochromatic columns
(such as $R_k$) followed by a multicoloured column (like $L$).
The columns $L$ and $R_k$ are shown in Fig. \ref{fig7}(a)
for the case
$K=3$.  Columns not relevant to the present discussion are not shown.
We label the original colour of $L$ as $C_0$ and the colour of $R_k$ as
$C_k$.  Various distances (which
are numerically equal to areas since the columns are of unit width) are
marked on the figure.
The strategy we will adopt is the following. We note that, as things
stand, the colour $C_k$ in all of $R_k$ also appears in $L$ and
hence, there must be some of the same colour in different columns for
the same $q$.  To correct this, we can swap an area, $Y_k$, of colour $C_0$
from $L$ into $R_k$ thus swapping the same area, $Y_k$, of $C_k$ back
into $L$.  We do this for all $k$. We then re-sort column $L$ so that
$Y_k$ lies immediately below $X_k$,
both these areas being of the same colour $C_k$ and so that colour $C_1$
is above colour $C_2$, etc.  This is shown in
Fig. \ref{fig7}(b) (the role of area $W$ will be explained later).
We choose $Y_K$ to be such that the proportion of
$C_k$ in $L$
relative to the height of $L$ is equal to the proportion of $C_0$ in
$R_k$ relative to the height of $R_k$.  This condition can be expressed
as
\begin{equation}\label{props}
  {X_k+Y_k\over \lambda'_L} =  {Y_k \over \lambda'_{R_k}}
\end{equation}
Furthermore, the area $Y_k$ of colour $C_0$ is placed in $R_k$ at the same
relative position of $R_k$ as the area $X_k+Y_k$ of colour $C_k$ has
been placed in column $L$ (see Fig. \ref{fig7}(b)).  This ensures that when the
columns are divided up as shown in Fig. \ref{fig6}, but with $Q$ large,
the pieces
$[q,i_{L}]$ in $L$ of colour $C_k$ will have a different colour from the
pieces $[q,i_{R_k}]$ in $R_k$ since the latter pieces will be of colour
$C_0$.

Having carried out this correcting procedure for these columns we see
that there is a problem.  Piece $S'$ which has been swapped back
somewhere to the left of $L$ is of colour $C_0$.  This piece may overlap
(in the sense of occupying some pieces with the same $q$) with the piece
$Y_1$, also of colour $C_0$,
which is in column $R_1$.  We can see $S'$ will not also overlap with the
pieces $Y_2,Y_3,\dots$ (also of colour $C_0$) in columns $R_2,R_3,\dots$
for the following reasons: (1) It is smaller than $X_1$ and hence can
only partly overlap with the piece $X_1+Y_1$ (of colour $C_1$) in column $L$ in
Fig. \ref{fig7}; (2) This piece of colour $C_1$ in column $L$ does not
overlap with the colour $C_0$ in columns $R_2, R_3,\dots$.  Hence, this
problem only
concerns piece $S'$ and column $R_1$.  Let us assume that piece $S'$ is in
column $L'$ and that the original colour of column $L'$ is
$C'_0$. We can use colour $C'_0$ to correct for piece $S'$ in $L'$
and $Y_1$ in
$R_1$ by essentially the same correcting procedure as before.  Thus, we
swap a piece of $C'_0$ and of area $W'$ from $L'$ to into $Y_1$ in $R_1$ and
at the same time swap a piece of colour $C_0$ and of area $W'$ from
$Y_1$ in $R_1$ into $L'$.  The areas $S'$ and $W'$ are collected
together at the top of column $L$. The size of area $W'$ is chosen to be
just such that $S'+W'$ (which is of colour $C_0$) no longer overlaps with
any of the colour $C_0$ in $R_1$.  The maximum size of area $W'$ is
given by
\begin{equation}
{S'+W'_{\max} \over \lambda'_{L'}} = {W_{\max} \over \lambda'_{R_1}}
\end{equation}
We could have $W'$ smaller than the value given by this equation since
it is possible that not all of $S'$ overlaps with $Y_1$ in $R_1$.
Hence,
\begin{equation}\label{wpmax}
W' \leq {S \lambda'_{R_1} \over \lambda'_{L'}-\lambda'_{R_1}}
\end{equation}

This correcting procedure is carried out in the following way.  First
the rightmost bunch of columns like $R_k$ and $L$ are selected.  Then
the directly swapped pieces are corrected.  And then the swapped back
pieces are corrected.  Then the next bunch of columns are
subject to the same correcting protocol until all the bunches have been
corrected.  If this procedure can be carried out
successfully then we will have solved this colouring problem.  The only
possible problem would be if we had to swap more of the colour $C_0$
from $L$ than there is in the column.  However, we can show that this will not
happen.  From the Fig. \ref{fig7}(a) we see that the area,
$z$, of the original colour $C_0$ in $L$ satisfies
\begin{equation}\label{higher}
\lambda'_L -\sum_{k=1}^K X_k-S = z \geq \lambda'_{R_1} \geq \lambda'_{R_k}
\end{equation}
The inequality must be satisfied since otherwise elements of
colour $C_1$ in row $L$ will be in the same row as elements of colour
$C_1$ in row $R_1$ but we proved in Sec. II. B. that this could not
happen with this
colouring of the diagram.  From (\ref{props},\ref{higher}) we obtain
\begin{equation}
Y_k={\lambda'_{R_k} \over \lambda'_L-\lambda'_{R_K}}X_k
\leq {\lambda'_{R_k} X_k\over \sum_{k'} X_{k'} +S}
\leq {z X_k\over \sum_{k'} X_{k'}+S}
\end{equation}
Since $S'$ had to be corrected for, $S$ must also be corrected for (if
$L$ and $R_k$ are taken to represent a general bunch of columns).  To
correct for $S$ will require an area $W$ of colour $C_0$ to be swapped into
a column $R''_1$ somewhere on the right.  By analogy with (\ref{wpmax})
we have
\begin{equation}\label{wmax}
W \leq {S \lambda'_{R''_1} \over \lambda'_{L}-\lambda'_{R''_1}}
  \leq {S z\over \sum_{k'} X_{k'} +S}
\end{equation}
where the second inequality follows since $\lambda'_{R''_1} \leq
\lambda_{R_1}\leq z$.  Hence,
\begin{equation}
\sum_k Y_k +W \leq z
\end{equation}
Which means that there is enough of the colour $C_0$ in column
$L$ to complete this colouring strategy.  Hence, the colouring problem
has been solved.

\subsection{Proof that Nielsen's bound cannot be beaten}

\begin{figure}
\resizebox{5in}{!}
{\includegraphics{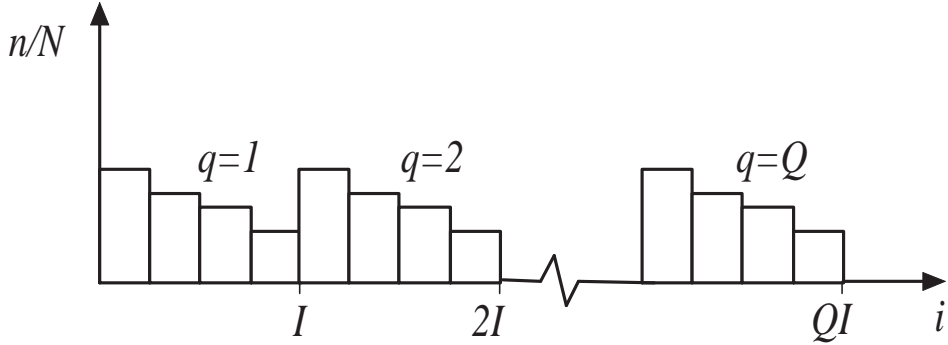}}
\caption{
The mini-step form consist of the pieces being arranged along the
$i$-axis starting with the $q=1$ pieces.
}\label{fig8}
\end{figure}

We need to prove that Nielsen's bound cannot be beaten.  This is
equivalent to proving that it is not possible to have net movement of
area down the area diagram.  The start state
corresponds to a step structure.  Each column in this can be divided up
into $Q$ pieces as shown in Fig. \ref{fig6}.  Next we collect together all the
pieces corresponding to a given $q$ and place them in order along the
$i$ axis as shown in Fig. \ref{fig8}.
Thus, going along the $i$ axis we have the
$q=1$ pieces, then the $q=2$ pieces and so on.  This can be accomplished
by Alice performing swap operations.  We will call this the
{\it mini-step form} for the area diagram.  (We are, of course, assuming that
system $A$ has a large enough Hilbert space to be able to do this.  If
this is not the case
then an additional ancilla could be introduced to effectively increase
the size of $A$'s Hilbert space.)  Let the state corresponding to this
diagram be $|\Psi_{\rm start}^{\rm mini}\rangle$.  Now we change from
the $(n,i)$ to
the $\{ n,j\}$ picture where $j$ is the $j$ in $|j\rangle_B$.  The
$\{ n,j\}$ element can only be moved up and down the diagram when Alice
performs local operations.  Hence, in the
$\{ n,j\}$ picture all the mini-steps will overlay each other so there will
be $Q$ elements at each position.
The height of the $j$th column will
be $N_j/Q$. Hence, if we define $A_{nj}=|{}_B\langle j|{}_S\langle
n|\Psi^{\rm mini}\rangle|^2$, then for the start state
$|\Psi_{\rm start}^{\rm mini}\rangle$ we have
\begin{equation}\label{initiala}
\sum_{n=1}^{n'} A_{nj}^{\rm start} = {n'Q\over N}
\end{equation}
for all $n'\leq N_j/Q$.  The factor $Q$ comes from the fact that there
are $Q$ sets of mini-steps overlaying each other in the $\{ n,j\}$
picture.

\begin{figure}
\resizebox{1.5in}{!}
{\includegraphics{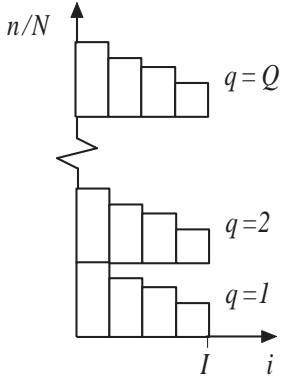}}
\caption{
This shows an intermediate state of the area diagram involved in proving that
Nielsen's bound cannot be beaten.
}\label{fig9}
\end{figure}

Now we go back to the $(n,i)$ picture and consider the target state
(\ref{targetg}).  We can apply the transformation
\begin{equation}
|l\rangle_S \rightarrow {1\over\sqrt{V_l}} \sum_{q\in W_l}
|q\rangle_S
\end{equation}
where $V_l=\mu_l Q$, $W_l$ is the set of integers from $(\sum_{r=0}^{l-1}
V_r)+1$ to $\sum_{r=0}^l V_r$ (we set $V_0=0$).  We will let
$Q\rightarrow\infty$.
Under this transformation (\ref{targetg}) becomes
\begin{equation}\label{targetf}
|\Psi''_{\rm target}\rangle={1\over\sqrt{Q}}\sum_{q\in W_l}
|q\rangle_S|\psi'\rangle^q
\end{equation}
where the superscript $q$ is equal to $l$ for $q\in W_l$.  This state
now consists of a number of terms $|q\rangle_S|\psi'\rangle^q$ each having the
same amplitude ${1\over\sqrt{Q}}$.  Each term,
$|q\rangle_S|\psi'\rangle^q$, can individually be put into the step form
\begin{equation}
|q\rangle_S|\psi'\rangle^q\rightarrow |\Psi',q\rangle=
{\sqrt{Q\over N}}\sum_i\sum_{n=(q-1)N/Q+1}^{qN/Q}
|n\rangle_S|i\rangle_A|i\rangle_B^q
\end{equation}
by applying transformations similar to (\ref{firsttrans}).
If this transformation is applied to all terms then the resulting area
diagram will consist of a series of mini-steps lined up vertically as
shown in Fig. \ref{fig9}.
Next, Alice applies swap operations to move these sets of mini-steps
so that they are lined up along the $i$ axis starting with the $q=1$ pieces
giving an area diagram in mini-step form (like in Fig. \ref{fig8}).  The
state becomes
\begin{equation}
|\Psi_{\rm target}^{\rm mini}\rangle=
{\sqrt{1\over N}}\sum_q\sum_j\sum_{n=1}^{N/Q}
|n\rangle_S|j+I(q-1)\rangle_A|j\rangle_B^q
\end{equation}
For this state we have
\begin{equation}\label{thetaarea}
|{}_B\langle \theta|{}_S\langle n|\Psi_{\rm target}^{\rm mini}\rangle|^2
\leq  {1\over N}\sum_q 1 ={Q\over N}
\end{equation}
for any normalised state $|\theta\rangle_B$.

The problem is to go from the start diagram in mini-step form to the
target diagram which is also in mini-step form.  If and only if we can
do this can we also go between the corresponding diagrams in standard
step form since Alice can transform reversibly between the two types of form
of the area diagram.
If there is to be net movement of area downwards in the mini-step form
then this must happen for at least one value of $j$.  Hence,
comparing with (\ref{initiala}), net downward movement of area implies
\begin{equation}\label{nbigarea}
\sum_{n=1}^{n'} A_{nj}^{\rm target} > {n'Q\over N}
\end{equation}
for at least one value of $j$ and $n'<N_j/Q$.
However, (\ref{thetaarea}) implies
\begin{equation}\label{nsmallarea}
\sum_{n=1}^{n'} A_{nj}^{\rm target}
=\sum_{n=1}^{n'} |{}_B\langle j|{}_S\langle n|\Psi_{\rm target}^{\rm
mini}\rangle|^2 \leq {n'Q\over N}
\end{equation}
which contradicts (\ref{nbigarea}) and hence there can be no
net movement of area downwards in the mini-step form.  The standard
step form area diagrams are simply elongated versions of one set of
mini-steps in the mini-step form, and hence, by the similarity of
these shapes, there can be no movement of area downwards in the standard
picture.  This proves Nielsen's bound (given algebraically in
(\ref{newsteps})).

\section{Conclusions}

In this paper a method of areas has been developed which enables us to
understand the manipulation of pure two-particle entanglement.
This approach has been used to find the most general way of transforming
a general two-particle pure state into maximally entangled states.
Certain results of Lo and Popescu were given geometric interpretations.
This method has also been used to prove Nielsen's theorem which pertains
to going from one two-particle pure state to another with certainty.
There remain a number of open problems relating to manipulation of
two-particle pure entanglement which it may be possible to solve using
the method of areas. Firstly, we could generalise Nielsen's theorem to
the problem where we go from one state to another but not necessarily
with certainty.  Secondly, we could consider the problem of going from
one state to a distribution of states.  The method may also
generalise to more than two particles (though it is not presently clear
how this generalisation will work).

\vspace{6mm}

{\bf Acknowledgements}

\vspace{6mm}

I am grateful to Anna Mahtani for suggesting a way to solve the second
colouring
problem and to Daniel Jonathan for comments relating to section 2.4 in
an earlier version.  I would also like to thank the Royal Society for
funding.

Note:  G. Vidal pointed out to the author that he had already solved
\cite{vidal} the first open problem mentioned in the conclusion.  The second
open problem
has been solved by D. Jonathan and M. Plenio \cite{jonple} in work done
independently of the present work.  They also use their
result to give a derivation of equation (\ref{emax}).

\vspace{6mm}

{\bf Appendix}

\vspace{6mm}

In this appendix we show that there can be no advantage if Alice makes a
non-maximal rather than a maximal measurement onto  $S$.
Assume that the state just before measurement is
\begin{equation}\label{gen}
\sum_l c_l |l\rangle_S|\Phi_l\rangle
\end{equation}
where $|\Phi_l\rangle$ is some state of system $AB$ and not necessarily
an $m$-state.
Imagine that the projective measurement is non-maximal and one of its
projectors is $|1\rangle_S\langle 1|+|2\rangle_S\langle 2|$.  In the
case of having the corresponding outcome, the resulting (unnormalised)
state will be
\begin{equation}\label{non}
c_1|1\rangle_S|\Phi_1\rangle+c_2|2\rangle_S|\Phi_2\rangle
\end{equation}
This could, for example, be an $m$-state if system $S$ is regarded as
being part of system $A$.  Rather than performing this non-maximal
measurement Alice could instead change her notation for the
$|i\rangle_A$ states such that, if $|i\rangle_A$ appears in the
expansions of $|\Phi_1\rangle$ and $|\Phi_2\rangle$ she writes
$|i\rangle_A$ as $|1,i\rangle_A$.
The remaining vectors $|i\rangle_A$ are relabelled as $|2,i\rangle_A$.
We are free to assume that the dimension of $A$ is big enough to do
this.  Then we write $|k,i\rangle_A=|k\rangle_{A'}|i\rangle_A$.
Now Alice performs the transformations
\begin{equation}
|1\rangle_S|1\rangle_{A'} \rightarrow |2\rangle_S|1\rangle_{A'}
\end{equation}
\begin{equation}
|2\rangle_S|1\rangle_{A'} \rightarrow |2\rangle_S|2\rangle_{A'}
\end{equation}
Under these transformations the first two terms in (\ref{gen}) become
\begin{equation}
|2\rangle_S(c_1|1\rangle_{A'}|\Phi_1\rangle+c_2|2\rangle_{A'}|\Phi_2\rangle)
\end{equation}
A maximal measurement will now give rise to a state with the same form
as the state as in (\ref{non}) for the outcome 2.
This trick can be repeated everywhere there is a degeneracy in the
original non-maximal measurement and a maximal measurement can then be
performed instead.  This maximal measurement will give rise to the same
distribution of the same states as the non-maximal measurement and so
there can be no advantage to performing non-maximal measurements.

\end{document}